%% file: main.tex
\documentclass{IEEEcsmag}

\usepackage[colorlinks,urlcolor=blue,linkcolor=blue,citecolor=blue]{hyperref}
\expandafter\def\expandafter\UrlBreaks\expandafter{\UrlBreaks\do\/\do\*\do\-\do\~\do\'\do\"\do\-}
\usepackage{upmath,color}

\usepackage{graphicx}
\usepackage{epstopdf}
\usepackage{epsfig}
\usepackage{alltt}
\usepackage{times}
\usepackage{caption}
\usepackage{tabularx}
\usepackage{pifont}

\usepackage{etoolbox}


\usepackage{todonotes}
\usepackage{comment}
\usepackage{xspace}

\let\oldFootnote\footnote
\newcommand\nextToken\relax

\renewcommand\footnote[1]{%
    \oldFootnote{#1}\futurelet\nextToken\isFootnote}

\newcommand\isFootnote{%
    \ifx\footnote\nextToken\textsuperscript{,}\fi}

\newcommand{\ie}{\emph{i.e.},\xspace}
\newcommand{\eg}{\emph{e.g.},\xspace}

\let\mycolor\color
\renewcommand{\mycolor}[2][]{}

\jvol{XX}
\jnum{XX}
\paper{8}
\jmonth{Month}
\jname{Publication Name}
\jtitle{Publication Title}
\pubyear{2021}

\begin{document}

\sptitle{Security \& Privacy Magazine}

\title{Verifiable Sustainability in Data Centers}

\author{Syed Rafiul Hussain}
\affil{Pennsylvania State University, hussain1@psu.edu}

\author{Patrick McDaniel}
\affil{University of Wisconsin--Madison, mcdaniel@cs.wisc.edu}

\author{Anshul Gandhi}
\affil{Stony Brook University, anshul@cs.stonybrook.edu}

\author{Kanad Ghose}
\affil{Binghamton University, ghose@binghamton.edu}

\author{Kartik Gopalan}
\affil{Binghamton University, kartik@binghamton.edu}

\author{Dongyoon Lee}
\affil{Stony Brook University, dongyoon@cs.stonybrook.edu}

\author{Yu David Liu}
\affil{Binghamton University, davidl@binghamton.edu}

\author{Zhenhua Liu}
\affil{Stony Brook University, zhenhua.liu@stonybrook.edu}

\author{Shuai Mu}
\affil{Stony Brook University, shuai@cs.stonybrook.edu}

\author{Erez Zadok}
\affil{Stony Brook University, erez.zadok@stonybrook.edu}

\markboth{DEPARTMENT}{DEPARTMENT}

\begin{abstract}%\looseness-1
Data centers have significant energy needs, both embodied and operational, affecting sustainability adversely.
The current techniques and tools for collecting, aggregating, and reporting verifiable sustainability data are
vulnerable to cyberattacks and misuse, requiring new security and privacy-preserving solutions.
This paper outlines security challenges and research directions for addressing these pressing requirements.
\end{abstract}

\maketitle

\input{intro3}
\input{systems}
\input{security-challenges}
\input{verifying}
\input{conc}

\section*{Acknowledgments}
The work reported in this paper has been supported by NSF under grants 2215017, % Penn State CNS Large
2214980, %  SBU - CNS Large
2046444, % Zhenhua CAREER
2215016, and % Kanad and David - CNS Large
2106263. % Erez-CNS Medium

\def\refname{REFERENCES}

\bibliographystyle{plain}
\bibliography{main}

\begin{IEEEbiography}{S. R. Hussain}{\,}
 is an Assistant Professor in the Computer Science and Engineering Department at
 Pennsylvania State University.
 His research interests include systems and network security, formal methods, and sustainability.
 Hussain received the Ph.D. degree in Computer Science from Purdue University, West Lafayette.
He is a Member of ACM and IEEE.  Contact him at hussain1@psu.edu.
\end{IEEEbiography}

\begin{IEEEbiography}{P. McDaniel}{\,} is
  the Tsun-Ming Shih Professor of Computer Sciences in the School of Computer,
  Data \& Information Sciences at the University of Wisconsin-Madison.
  McDaniel's research focuses on a wide range of topics in computer and network
  security and technical public policy, with interests in mobile device security,
  the security of machine learning, systems, program analysis for security,
  sustainability and election systems. McDaniel is a Fellow of IEEE, ACM and AAAS,
  a recipient of the SIGOPS Hall of Fame Award and SIGSAC Outstanding Innovation Award,
  and the director of the NSF Frontier Center for Trustworthy Machine Learning.
  He also served as the program manager and lead scientist for the Army Research Laboratory's
  Cyber-Security Collaborative Research Alliance from 2013 to 2018.
  Prior to joining Wisconsin in 2022, he was the William L. Weiss Professor
  of Information and Communications Technology and Director of the Institute
  for Networking and Security Research at Pennsylvania State University. Contact him
  at mcdaniel@cs.wisc.edu.
\end{IEEEbiography}

\begin{IEEEbiography}{A. Gandhi} {\,} is an Associate Professor in the Computer Science Department at Stony Brook
  University.
  His research interests include performance analysis, modeling, and evaluation; distributed systems; and sustainability.
  Gandhi received the Ph.D. degree in Computer Science from Carnegie Mellon University.
  He is a Senior Member of ACM and a Senior Member of IEEE.
  Contact him at anshul@cs.stonybrook.edu.
\end{IEEEbiography}

\begin{IEEEbiography}{K. Ghose} {\,} is a SUNY Distinguished Professor of Computer Science at SUNY Binghamton (Binghamton University).
  His research interests include energy-aware systems at all scales, processor microarchitectures and hardware security.
  Ghose received the Ph.D. degree in Computer Science from Iowa State University.  He is a Member of ACM and IEEE.
  Contact him at ghose@binghamton.edu.
\end{IEEEbiography}

\begin{IEEEbiography}{K. Gopalan} {\,} is a Professor in the Computer Science Department at Binghamton University.
His research interests are in computer systems including virtualization, security, operating systems, networks, and sustainability.
He received his Ph.D. degree in Computer Science from Stony Brook University.
He is a senior member of IEEE and a member of ACM.
Contact him at kartik@binghamton.edu.
\end{IEEEbiography}

\begin{IEEEbiography}{D. Lee} {\,} is an Assistant Professor in the Computer Science Department at Stony Brook University.
His research interests include compilers, operating systems, computer architecture, security, and sustainability.
Lee received the Ph.D. degree in Computer Science Engineering from Michigan University, Ann Arbor.
He is a Member of ACM and IEEE.
Contact him at dongyoon@cs.stonybrook.edu.
\end{IEEEbiography}

\begin{IEEEbiography}{Y. D. Liu} {\,} is a Professor in the Computer Science Department at Binghamton
  University.
  His research interests include programming languages; software engineering; formal methods for security; sustainable and energy-aware applications and systems.
  Liu received the Ph.D. degree in Computer Science from the Johns Hopkins University.
  He is a Member of ACM.
  Contact him at davidl@binghamton.edu.
\end{IEEEbiography}

\begin{IEEEbiography}{Z. Liu} {\,} is an Associate Professor in the Department of Applied Mathematics and Statistics and Department of Computer Science at Stony Brook University. His research interests include optimization; machine learning; big data systems;  sustainable and energy-aware applications and systems. Liu received the Ph.D. degree in Computer Science from California Institute of Technology. He is a Member of ACM. Contact him at zhenhua.liu@stonybrook.edu.
\end{IEEEbiography}

\begin{IEEEbiography}{S. Mu} {\,} is an Assistant Professor in the
Department of Computer Science at Stony Brook University.
His research interests include distributed systems, multi-core systems, and sustainable systems.
Mu received the Ph.D. degree in Computer Science from Tsinghua University.
He is a Member of ACM. Contact him at shuai@cs.stonybrook.edu.
\end{IEEEbiography}

\begin{IEEEbiography}{E. Zadok} {\,} is a Professor at Stony Brook
  University.
  His research interests include computer systems, storage systems,
  security, performance optimizations, and sustainability.
  Zadok received the Ph.D. degree  in Computer Science from Columbia
  University.
  He is a Senior Member of the IEEE Computer Society, an ACM Distinguished
  Member, and a member of USENIX.  Contact him at Erez.Zadok@stonybrook.edu.
\end{IEEEbiography}

\end{document}

%% file: intro3.tex
% \section{Introduction}
% \label{s:intro}

\chapteri{S}ustainability is the practice of performing human
activities in ways that do not leave lasting harmful
effects. %~\cite{un22}.
Unfortunately, the harm to the planet is clearly growing\footnote{\href{https://www.computerworld.com/article/3431148/why-data-centres-are-the-new-frontier-in-the-fight-against-climate-change.html}
{https://www.computerworld.com/article/3431148/why-data-centres-are-the-new-frontier-in-the-fight-against-climate-change.html}}\footnote{Environmental footprint of data centers: Md Abu Bakar Siddik et al., 2021 Environ. Res. Lett.},
whether the effects are direct (\eg emissions
caused by transportation, farming, or manufacturing) or indirect
(\eg
carbon emissions due to electricity consumed by data centers
and even the energy and materials used for
manufacturing servers and other devices).
Humans as a species have understood
that sustainability is important to both future generations and the
global quality of life. Yet, we have had only sporadic and uneven
adoption of sustainable practices, and up to 98\% of sustainability
initiatives fail to meet their goals\footnote{\href{https://www.bain.com/insights/achieving-breakthrough-results-in-sustainability}{https://www.bain.com/insights/achieving-breakthrough-results-in-sustainability}}.
%~\cite{dst16}.
The impacts of a
lack of sustainability have led to---among many other
factors---climate change, widespread pollution of the oceans, sea
bottom desertification,
acidification of land and water, ozone loss, desertization, and loss
of biodiversity. Failure to address this lack of sustainability now
will create long-term problems for future generations. %~\cite{ipcc22}.

Today, achieving the goals of sustainability requires the honest, best
efforts of humans and an apparatus to measure aspects of the system
under regulation.  Yet, those efforts often fail when bad actors
bypass or cheat sustainability systems.  For example, the car company
Volkswagen installed emissions software on roughly 11 million cars
worldwide that misled the Environmental Protection Agency (EPA) about
emissions when under test\footnote{\href{https://www.bbc.com/news/business-34324772}{https://www.bbc.com/news/business-34324772}}.
%~\cite{vwscandal2015}.
Volkswagen was eventually caught, fined
billions of dollars, and required to recall vehicles and pay financial
settlements---but only \emph{after} the vehicles had polluted for
nearly a decade.

One area with unprecedented impact on our world is the use of
computation and in particular data centers.
With the alarming rise of computation and the pervasive use of
artificial intelligence (\eg ChatGPT)\footnote{\href{https://www.nnlabs.org/power-requirements-of-large-language-models/}{https://www.nnlabs.org/power-requirements-of-large-language-models/}}, %~\cite{llm-energy},
data centers pose many negative impacts
on the environment caused by energy use, hardware
manufacturing and disposal, building maintenance, water usage 
and other factors. 
Indeed, a
recent study showed that over 2--4\%\footnote{\href{https://www.iea.org/reports/data-centres-and-data-transmission-networks}{https://www.iea.org/reports/data-centres-and-data-transmission-networks}}
of all energy used worldwide was by 
data centers~\cite{masanet20}. %, iea}.
The current practice of reporting
sustainability information in data centers is, however, mired with
``greenwashing,'' where the true carbon footprint of a data center
is artificially reduced via the purchase of energy or certificates from green generation
sources\footnote{\label{drec}\href{https://drecs.org/}{https://drecs.org/}}
%~\cite{drec-initiative}
or by paying other entities to be
sustainable.
This signifies a lack of transparency and accountability
that hinder efforts to address and mitigate the environmental
consequences associated with data centers.  Such issues are
pervasive as they extend beyond data centers and permeate various industries,
including food, manufacturing, and telecommunication systems.

The lack of accountability and transparency to address sustainability
is primarily rooted in the absence of \emph{complete} and \emph{verifiable}
sustainability data and metrics\footnote{\href{https://www2.deloitte.com/us/en/insights/industry/retail-distribution/accountable-sustainability-consumer-industry.html}{https://www2.deloitte.com/us/en/insights/industry/retail-distribution/accountable-sustainability-consumer-industry.html}}.
%\footnote{\href{https://www.imf.org/en/Blogs/Articles/2021/05/13/how-strengthening-standards-for-data-and-disclosure-can-make-for-a-greener-future}{https://www.imf.org/en/Blogs/Articles/2021/05/13/how-strengthening-standards-for-data-and-disclosure-can-make-for-a-greener-future}}
%\href{https://www.brookings.edu/blog/future-development/2021/09/28/the-risks-of-us-eu-divergence-on-corporate-sustainability-disclosure/}{https://www.brookings.edu/blog/future-development/2021/09/28/the-risks-of-us-eu-divergence-on-corporate-sustainability-disclosure/}}.
%~\cite{data-disclosure-greener-future,
%accountable-sustainability, corporate-sustainability-disclosure}.
Comprehensive and fine-grained sustainability metrics
%~\cite{nsfgdi}
are critical to identify performance bottlenecks (\eg the impact of an
application's code or library on sustainability), diagnose security
issues, detect anomalous sustainability activities, provide reliable
audit trail of carbon consumption, ensure accurate and precise
accountability and compliance benefits (\eg
accurately identify entities who made changes or performed certain
actions), and optimize system performance\footnote{\href{https://www.nsf.gov/pubs/2020/nsf20594/nsf20594.htm}{https://www.nsf.gov/pubs/2020/nsf20594/nsf20594.htm}}.
% ~\cite{nsfgdi}.
% ~\cite{bashir21-sustainableclouds,
% chasing-carbon-udit2022, nsfgdi}.
Therefore, a necessary first step for any sustainable computing approach is the
ability to measure comprehensive sustainability metrics or cost
functions from all possible sources of carbon consumption and energy
spent in the entire life-cycle of the computing equipment: production,
delivery, and disposal; these are referred to as ``embodied energy.''
However, it has been
found that it is difficult to determine
accurate sustainability metrics because the sources are too many,
untrustworthy, disconnected, or incompatible. Further,
there is no way to combine the data in a meaningful way that will not
compromise the privacy of users or service
providers\footnote{\href{https://hbr.org/2021/05/overselling-sustainability-reporting}{https://hbr.org/2021/05/overselling-sustainability-reporting}}.
%~\cite{overselling-sustainability-reporting}.
For example,
there are dozens of different ways to calculate data on global data
center energy consumption based on public and private data---each resulting
in an assessment that is often contradictory with others\footnote{\href{https://energyinnovation.org/2020/03/17/how-much-energy-do-data-centers-really-use/}{https://energyinnovation.org/2020/03/17/how-much-energy-do-data-centers-really-use/}}.
%~\cite{eip20}.
Hence, we have at best a vague idea of the impact that, for example,
data centers have on our environment.
Even when attempts are made to collect
and combine sustainability metrics from disparate sources,
privacy concerns, exposure of sensitive users' data or service
providers' proprietary algorithms are often ignored, resulting in poor
incentives for users or service providers to opt for accountable
sustainability systems.
Researchers and organizations
trying to understand and create sustainable systems often refer to the
\emph{sustainability data gap}.
The inability to collect and verify
accurate, complete, and timely data on the environment in a
privacy-preserving fashion is slowing, and in some cases prohibiting,
the adoption of sustainable systems and practices.
To make matters worse, market forces and
human greed, as we observed earlier, often work against
the goals of sustainability.

In the context of data centers, which is the primary focus of this paper,
the infrastructures used to measure and maintain \emph{operational
sustainability}
(\ie environmental footprints transpired within a data center) are
inherently \emph{adversarial}:
because users of technology (\eg data center users)
have an incentive to cheat, the apparatus must strive to ensure that systems continue to
function correctly in the face of actors attempting to thwart the
collection of
sensitive sustainability footprint and the enforcement of corresponding
security and privacy policies. 
Hence, it is imperative that the environmental footprint
caused by data center operations can be verified by
interested third parties (\eg the EPA\footnote{\href{https://www3.epa.gov/carbon-footprint-calculator/tool/definitions/co2e.html}{https://www3.epa.gov/carbon-footprint-calculator/tool/definitions/co2e.html}},
%~\cite{co2e_epa},
citizen scientists, and the public).

This article, therefore, identifies the security issues in the
sustainability data pipeline comprising of data \emph{collection},
\emph{storage}, \emph{aggregation} (or other processing),
\emph{reporting} and \emph{use} \emph{in situ}.
More specifically, we examine threat landscapes and a wide
range of security challenges to build verifiable sustainability within
data centers, highlighting the urgent need to address these
threats.
Furthermore, we explore a variety of promising research directions
that will yield novel and practical solutions to combat these security
challenges in sustainable data centers and mitigate the risks
associated with such threat landscapes.
Some of our proposed security challenges and solutions also apply to
other industry segments: manufacturing, airlines and transportation,
industrial-scale farming, and more.

%%%%%%%%%%%%%%%%%%%%%%%%%%%%%%%%%%%%%%%%%%%%%%%%%%%%%%%%%%%%%%%%%%%%%%%%%%%%%%
%% For Emacs:
% Local variables:
% fill-column: 70
% End:
%%%%%%%%%%%%%%%%%%%%%%%%%%%%%%%%%%%%%%%%%%%%%%%%%%%%%%%%%%%%%%%%%%%%%%%%%%%%%%
%% For vim:
% vim:textwidth=70
%%%%%%%%%%%%%%%%%%%%%%%%%%%%%%%%%%%%%%%%%%%%%%%%%%%%%%%%%%%%%%%%%%%%%%%%%%%%%%
% LocalWords:  PSU acidification desertization IPCC cybersecurity geo
% LocalWords:  incentivized ustainability Abu Bakar Siddik Environ
% LocalWords:  Lett ChatGPT greenwashing

%% file: systems.tex
\section{Sustainable Systems and Focus on Data Centers}
\label{s:systems}

There are several systems (or industries) whose unsustainable
operations pose a grave threat to the environment.
For example, sustainability concerns are important across a wide
industry segment such as livestock farming, automobiles, airlines,
manufacturing, energy generation, transportation, as well as
infrastructure construction and management (\eg those applicable
to buildings and roadways).
Data centers are particularly significant due to their substantial
energy consumption and environmental impact.
Moreover, data centers
play a vital role in supporting many industries and services that rely
on digital infrastructure, making their sustainability practices even
more critical.
Operations within data centers already contribute significantly
to the global carbon footprint~\cite{ren12}.
%, moghaddam11}.
%
The rise in popularity of resource-intensive Big Data, AI, crypto-currency, and
Machine-Learning workloads is poised to make data center operations
even more unsustainable\footnote{\href{https://datacenterfrontier.com/the-bitcoin-energy-debate-lessons-from-the-data-center-industry}{https://datacenterfrontier.com/the-bitcoin-energy-debate-lessons-from-the-data-center-industry}}.
% ~\cite{kaffes20, asyabi20, masanet20,
%   crypto-sustainability, bitcoin-sustainability}.
%
Estimates suggest
that data centers are already responsible for about 2--4\% of the total
greenhouse emissions; that is equivalent to the emissions of the
entire airline industry\footnote{\href{https://www.climateneutralgroup.com/en/news/carbon-emissions-of-data-center}{https://www.climateneutralgroup.com/en/news/carbon-emissions-of-data-center}}
%~\cite{carbon-dc}.
Worse, this figure for data
centers is soon expected to increase to 5--7\%
with the emergence of Large Language Models (LLMs) such as
GPT-4 %~\cite{zeus},
%, llm-energy, ai-energy},
and applications based
on LLMs, imposing a heavier toll on the environment\footnote{\href{https://semiengineering.com/ai-power-consumption-exploding/}{https://semiengineering.com/ai-power-consumption-exploding/}}.

\mycolor{blue}
This paper, therefore, specifically focuses on sustainability
in data centers. Even though 
data centers' energy consumption can be significant, their efficiency 
through shared resources---the ability to integrate renewable energy and
optimize computing power---can potentially reduce the overall footprint compared to distributed on-premise solutions, such as edge servers or 
private cloud.
Also, the sheer volume of computing resources to process a wide range of data,
optimized cooling systems, and advanced energy consumption equipment
in data centers enable us to obtain a comprehensive view of the environmental
impact as opposed to their on-site alternatives. 
This could be achieved by a fine-grained approach to measuring energy consumption/carbon,
which requires security primitives, as it could become a criterion for optimization,
fine-grained diagnosis and decision making similar to financial cost in the long term.
This effort is also necessary to complete net zero and aligns with the aspirations of the
largest IT companies\footnote{\href{https://www.datacenterdynamics.com/en/news/microsoft-and-20-others-call-for-carbon-accounting-standards/}{https://www.datacenterdynamics.com/en/news/microsoft-and-20-others-call-for-carbon-accounting-standards/}}.
Moreover, such transparent monitoring and verifiable audit of energy consumption will help data centers
and service providers establish trust with customers, investors, and regulators.
Therefore, in this paper, we specifically focus on sustainability
in data centers.
\mycolor{black}

Existing practices in data centers on reporting or advertising sustainability data are often fraught with greenwashing; as a result, the true carbon
usage of a data center is hidden. 
%~\cite{drec-initiative}.
Similar greenwashing practices have also been observed in other sectors including autonomous vehicles\footnote{\href{https://www.euronews.com/green/2023/01/16/driverless-cars-the-dark-side-of-autonomous-vehicles-that-no-ones-talking-about}{https://www.euronews.com/green/2023/01/16/driverless-cars-the-dark-side-of-autonomous-vehicles-that-no-ones-talking-about}}
%~\cite{autonomous-sustainability}
and telecommunication industries\footnote{\href{https://www.euronews.com/green/2023/01/16/driverless-cars-the-dark-side-of-autonomous-vehicles-that-no-ones-talking-about}{https://www.euronews.com/green/2023/01/16/driverless-cars-the-dark-side-of-autonomous-vehicles-that-no-ones-talking-about}}.
%~\cite{5g-impact-on-environment}.
%
Such deceptive approaches
undermine the transparency and credibility of sustainability claims,
making it difficult for stakeholders to make informed decisions.
The European Union's Corporate Sustainability Reporting Directive
(CSRD) 
%\footnote{\href{https://www.europarl.europa.eu/legislative-train/carriage/review-of-the-non-financial-reporting-directive/report?sid=6501}{https://www.europarl.europa.eu/legislative-train/carriage/review-of-the-non-financial-reporting-directive/report?sid=6501}}
%~\cite{EUsusdir}
mandates that by 2024,
corporations have to report non-financial
sustainability information precisely and clearly; this
will also apply to data center operators within the EU.
There is some consensus among data center operators on
reporting data center sustainability information and metrics
accurately, at least within the EU and the Asia-Pacific
region
\footnote{\href{https://www.datacenterdynamics.com/en/marketwatch/reading-the-runes-eu-data-center-regulations-are-coming-sooner-than-you-think/}{https://www.datacenterdynamics.com/en/marketwatch/reading-the-runes-eu-data-center-regulations-are-coming-sooner-than-you-think/}}.
%~\cite{EU, EUAP}.
%
In the U.S., we also see the beginnings of directives similar\footnote{\href{https://www.sustainability.gov/pdfs/federal-sustainability-plan.pdf}{https://www.sustainability.gov/pdfs/federal-sustainability-plan.pdf}} to the
EUs,
%~\cite{fedplan},
but details are still emerging.

\mycolor{blue}
Coarse-grained accounting may be possible using recent carbon measurement prototypes,
such as Power API\footnote{\href{https://powerapi.org/}{https://powerapi.org/}},
Kepler\footnote{\href{https://sustainable-computing.io/}{https://sustainable-computing.io/}}, and Scaphandre\footnote{\href{https://github.com/hubblo-org/scaphandre}{https://github.com/hubblo-org/scaphandre}};
however,
they are still in their infancy and are not designed to provide any support for
the verifiability of the generated carbon footprints by regulatory agencies or to
ensure the privacy of users' sustainability data.
The lack of such security and privacy guarantees,
as outlined in Section~\ref{sec:security-challenges},
can be exploited by malicious entities (\eg data center and service
providers) to bypass carbon compliance, evade taxes, induce financial loss to rival companies,
cause over/under-billing to customers, steal and exfiltrate sensitive users' data and proprietary
models, and contribute to the environmental hazards.
Hence, security measures are indispensable for independent
audits to provide an objective and verifiable assessment of data centers'
sustainability claims, preventing greenwashing and ensuring accurate reporting of carbon emissions.
Also, this transparency and accountability builds trust with
stakeholders like customers, investors, and the general
public who are increasingly concerned about the environmental impact of data centers.

\mycolor{black}

\mycolor{blue}
\section{Data Center Architectures}
Data centers can be of different types based
on size, purpose, and the services they offer.
For instance, enterprise data centers are operated
by individual organizations to manage and store their
own data and IT infrastructure, whereas edge data centers
are smaller facilities located closer to end-users to reduce
latency and enhance the performance of
edge computing applications.
This paper, however, primarily focuses on
co-located, hyperscale, distributed or cloud data centers.
%(e.g., Google Cloud, Amazon AWS, and Microsoft Azure)
Co-located data centers (\eg Equinix, Digital Realty)
provide space, power, and cooling for servers
owned by different organizations (or \emph{tenants}),
promoting the sharing of the facility resources and physical infrastructures.
Hyperscale and cloud data centers such as Google, Amazon, and Microsoft,
deliver services like Infrastructure as a Service (IaaS),
Platform as a Service (PaaS), and Software as a Service (SaaS)
over the internet, and can handle massive amounts of data and traffic. 
IaaS providers 
abstract and virtualize the underlying physical IT infrastructure  
and create isolated virtual environments, thus enabling end-users and
customers to run applications, store data, and utilize physical
resources provided by data center providers. 
PaaS providers offer platforms (\eg development tools, middleware,
database and deployment services) as a service to SaaS providers to
streamline application development, deployment, and management,
facilitating faster and more efficient processes.  SaaS providers grant
users with instant access to software applications, data storage,
access control, APIs, and integration, eliminating the need for
businesses to invest in and maintain hardware and software, thereby
reducing overall IT infrastructure costs.
\mycolor{black}

%%%%%%%%%%%%%%%%%%%%%%%%%%%%%%%%%%%%%%%%%%%%%%%%%%%%%%%%%%%%%%%%%%%%%%%%%%%%%%
%% For Emacs:
% Local variables:
% fill-column: 70
% End:
%%%%%%%%%%%%%%%%%%%%%%%%%%%%%%%%%%%%%%%%%%%%%%%%%%%%%%%%%%%%%%%%%%%%%%%%%%%%%%
%% For vim:
% vim:textwidth=70
%%%%%%%%%%%%%%%%%%%%%%%%%%%%%%%%%%%%%%%%%%%%%%%%%%%%%%%%%%%%%%%%%%%%%%%%%%%%%%
% LocalWords:  PSU KANAD greenwashing CSRD EUs Scaphandre Equinix
% LocalWords:  IaaS PaaS

%% file: security-challenges.tex
%\section{Security Challenges of Sustainable Systems}

\section{Why is Sustainability a Security Problem?}
\label{sec:security-challenges}

Ensuring the accuracy and credibility of sustainability
metrics, as well as empowering audits by regulatory agencies,
require guaranteeing the trustworthiness
and comprehensiveness of not only the carbon footprints of data center
equipment but also the embodied energy throughout the entire life-cycle
of computing equipment.
Although some  external information---such as that for
renewable energy, energy credits, or supplied water---can be
authenticated via trusted third parties,
%~\cite{co2e_epa, iea},
sustainability metrics in data centers require the
authenticity, confidentiality, integrity, and availability of data
collected, processed, stored, and used locally within a
data center~\cite{gandhi2022metrics}.
However, unlike traditional cloud computing systems
where the focus is primarily on security and privacy of user applications
and data,  %~\cite{carlin2013cloud}, %
%, zissis2012addressing, chen2010s},
collecting and measuring data center activities that impact humans and
the environment in a verifiable and privacy-preserving manner
presents a diverse set of new security challenges.
Most of these challenges are primarily based on sustainability data,
reliability of equipment, and cleanliness of energy sources---across
both the digital and physical worlds.
Unfortunately, no prior research has investigated the threat landscape of
sustainable data centers, nor attempted to provide any techniques or tools
that directly allow authentication of operational sustainability
metrics induced within a data
center to preserve the privacy of users' or operators'
sustainability data.
\mycolor{blue}
Also, we note that a key distinction with sustainable data, as opposed
to regular data in cloud infrastructure, is that it is generated
independently of user intent, resulting in reduced trust guarantees.
The critical factor here is to prevent users from misrepresenting
their emissions.  Therefore, this data must be generated, collected,
and aggregated in a manner that is tamper-resistant, akin to a
physical value, ensuring that it is nearly impossible for anyone to
manipulate and does not expose any sensitive information about
users.  In a nutshell, this threat model is different from most common
data as the trust has to be minimal.
\mycolor{black}
It is thus imperative to ensure the security of (i) data collection
processes, (ii) the process of generating verifiable, easily auditable
sustainability metrics, and (iii) the storage of all pertinent
information.
Hence, while being indispensable for protecting the environment and our
planet, we have found and argue that the current sustainability
practices---through self-reporting, best-effort measurement, and
anything less than complete verifiable control of
sustainability---will fail.

\mycolor{blue}
\subsection{Threat Models in Data Centers}
\label{subsec:threat_model}

The trust assumptions and threat models for sustainable
data centers may vary widely based on data center type (\eg
multi-tenant and hyperscale vs.\ enterprise data centers), service models offered by
the data center or the tenants, and any other specific requirements.
In general,
the threat models for a co-located, hyperscale, or cloud data center's
sustainability can be primarily derived with respect to four entities:
(a) data center providers, (b) tenants or service providers,
(c) users, and (d) third-party observers (\eg regulatory
agencies) leading to the following adversarial capabilities.
($\mathcal{A}_1$) Here, data centers provide misleading or
false sustainability data to attract end-users or third-party
service providers.
($\mathcal{A}_2$) Data centers or tenants
providing IaaS, PaaS, and SaaS, often
characterized as \emph{honest but curious}, may attempt to
learn the proprietary or sensitive data of their users and
exfiltrate it to third parties.
($\mathcal{A}_3$) Data center or service providers have access to their users or tenant's sustainability data and can be inherently malicious to exploit this information to harm the users or learn proprietary information that would benefit competitors or harm their tenants/consumers.
($\mathcal{A}_4$) Tenants (\ie service providers), on the other hand,
can also subvert the security and privacy of the other co-located tenants' resources and the facilities provided to them. 
($\mathcal{A}_5$) To make matters worse, resources (hardware and software) served
by tenant (\eg IaaS, PaaS, or SaaS) within a data center can also
be compromised and controlled by external attackers who are nation-states or
rival organizations offering similar services.
This is possible
due to system/service misconfigurations, insecure communication protocols
inadequate access controls and isolation of shared and physical resources,
and vulnerabilities in the hardware, software, or other components of the service providers' supply chains.
For example, benign and unsuspecting data center providers
often use virtual machines (VMs) or containers created by IaaS providers
that are loaded with backdoors or malware illegitimately reading/writing sensitive
carbon footprint data.

($\mathcal{A}_6$) The other key entities in data centers (\ie users or customers of a tenant)
can also be considered malicious.  This is because a
user's job (\eg a process) running in a data center
may attempt to gain unauthorized access to read or modify
other jobs' code and data and thus affect the
sustainability data produced by other jobs. 
For example, a malicious process of an end user may
add unaccounted read/write operations~\cite{graphene_sgx_atc17}
%, glamdring}
to users' jobs which
can inflate users' carbon footprints, leading to overbilling the
victim customers. 
Such carbon footprint inflation can also be achieved by violating the integrity
of the sustainability metrics (\eg code or
data)~\cite{graphene_sgx_atc17}
%, glamdring}
or by manipulating the system traces and logs---the evidence trail of carbon
consumption
%~\cite{sgx_log_security_asiaccs17}
by the compromised VMs or
malicious processes in data centers.
Similarly, compromised data center providers may exploit the same and
use similar malicious processes to report false
carbon footprints to the regulators
%~\cite{sgx_use_based_privacy_wpes18}
to evade high carbon taxes or regulations~\cite{graphene_sgx_atc17}. 
Users may also try to launch attacks (\eg DoS)
against other users or the tenant who owns that service, another tenant or its users
in the same data center.  Users may also strive to obtain higher
levels of service than they are allocated, and thus mislead the service
providers about the user's carbon usage.
Various surfaces can be utilized by users to
attack the tenant, including the hypervisor, VMs, APIs and web services.

Last but not least, third-party observers (\eg regulatory agencies)
are tasked with verifying the footprint reported by the data center and service
providers in the process of executing policy or oversight (\eg by
comparing sustainability costs reported by cloud operators, users, and
utilities).  ($\mathcal{A}_7$) But even these observers
may be honest but curious, government or law enforcement agencies performing
surveillance,
or untrusted as they could collude with others to mislead reporting, may have rogue
insider elements within the data center, and may even be under political or other
pressure to ``fudge'' or misrepresent the data.
\mycolor{black}

\mycolor{blue}
\noindent\textbf{Differences with other systems}. Although there are some similarities between data centers and IoT and enterprise IT systems regarding data collection, verifiability and storage, the key distinction lies in the threat model between these systems.  
For instance, in most IoT settings, 
users, being the owners of their homes and devices, do not tamper with the devices to generate false data. The users also generally trust the 
trigger-action platforms capable of storing sensor data as those
platforms are the key enablers of automation. 
In most cases, the
third-party smart apps (\ie trigger-action rules) or external attackers 
are untrusted as they are the primary attack vectors.  Another
notable distinction with the sustainability data in data centers, as
compared to regular IoT data, is that the sustainability footprints
recorded by physical and virtual infrastructures (\eg power
generators, cooling systems, virtual machines, and hypervisors) are
shared across mutually untrusted stakeholders.  This gives rise to
privacy concerns, which inherently differ from those in IoT or enterprise IT  systems where multiple users share the same physical environment (\eg
smart home and building) are mutually trusted and hence one user's 
IoT activities are not considered sensitive/private to another user in the
same home/building.
\mycolor{black}

\input{table-challenges}
%%%%%%%%%%%%%%%%%%%%%%%%%%%%%%%%%%%%%%%%%%%%%%%%%%%%%%%%%%%%%%%%%%%%%%

\mycolor{blue}
\subsection{Security Challenges for Sustainability}
\label{subsec:new-security-challenges}
Due to the complex design of data centers and intricate interactions
among their stakeholders, it is necessary to characterize and
address diverse security threats on the sustainability pursuit of data centers. 
Next, we discuss some critical security
challenges for a data center aiming for sustainability,
summarized in Table~\ref{tab:sustainability_threats}.
Note that the nature of threats will be different for different
sustainable systems (\eg transportation, manufacturing) based on trust
assumptions.

\noindent \ding{113} \textbf{Evasive carbon offset techniques (C1).}
Data centers and large corporations often trade a known amount of carbon emissions with an
uncertain amount of emission reductions to claim carbon neutrality
(\eg by investing in forestation
elsewhere). %~\cite{tesla_carbon_offset}.
This practice, also called
carbon crediting or climate crediting, has been in place for decades.  It
is often exploited by large corporations as it is extremely difficult,
if not impossible, to track and verify if the amount of emissions
balances out the amount of reductions.
%~\cite{scam_carbon_offset}.
%~\cite{junk_carbon_offset, myth_carbon_offset, scam_carbon_offset}.
Often, Renewable Energy
Credits (RECs) are used to offset the carbon footprint of a data
center via the purchase of energy credits from a green energy
generator. %~\cite{drec-initiative}.
Similarly, Power Purchase
Agreements (PPAs)\footnote{\href{https://sourcingjournal.com/topics/sustainability/apparel-sustainability-unethical-practices-73619/}{https://sourcingjournal.com/topics/sustainability/apparel-sustainability-unethical-practices-73619/}}
%~\cite{ppa}
are used to have the data center operator
finance the installation of a green energy-producing farm, run, owned
and managed by an independent party, to provide green energy to the
data center over a long-term period covered under the PPA.  For both
RECs and PPAs, the authenticity of green energy is, however, often kept
out of sight of the users.  Therefore, the lack of authentication, accountability, and transparency
enables corporations ($\mathcal{A}_1$) to make false claims about the energy source,
while appearing in public to support sustainability efforts.

\noindent \ding{113} \textbf{Lack of integrity of carbon emission
  sources (C2).}
According to the threat model $\mathcal{A}_3$, sensors and devices (\eg PDUs) 
used for tracking
sustainability data can be tampered with by their owners, \ie
untrusted data centers, IaaS providers, or physically co-located $\mathcal{A}_4$ tenants to either
misreport to the regulatory agencies or overcharge the customers.
Such false reporting by $\mathcal{A}_4$ tenants can cause the data center operator to re-adjust
resource allocations/scheduling unnecessarily to adversely affect the data center's
sustainability footprint.
In a similar vein, these sensors and devices
can also become compromised by external attackers ($\mathcal{A}_5$) due to
unintentional vulnerabilities or intended backdoors in their hardware,
firmware, and software\footnote{\href{https://therecord.media/cisa-claroty-highlight-severe-vulnerabilities-in-popular-power-distribution-unit-product}{https://therecord.media/cisa-claroty-highlight-severe-vulnerabilities-in-popular-power-distribution-unit-product}}.
%~\cite{pdu-vulnerabilities}.
As a result, by
taking control of those sensors and devices, attackers may violate the
authenticity and forge carbon footprints to cause over/under-billing to customers by
forging/manipulating carbon consumption records.
Attackers may also generate false sustainability data or manipulate the cooling system
to disrupt sustainability operations\footnote{\href{https://www.datacenterknowledge.com/security/physical-infrastructure-cybersecurity-growing-problem-data-centers}{https://www.datacenterknowledge.com/security/physical-infrastructure-cybersecurity-growing-problem-data-centers}}.
Similar kinds of sustainability
data-forgery attacks can also be carried out if there are vulnerabilities in the
communication protocols (\eg lack of authentication and replay protection) between sensors and the
sustainability data aggregators gleaning carbon footprints from
multiple such sensors.
Due to such malicious actions,
additional water and electricity would be required to
cool the targeted data center, resulting in an increased
carbon footprint, higher operational costs, and disruption of
sustainability efforts.

\noindent \ding{113} \textbf{Inadequate access control and information
flow control (C3).}
While resource sharing in data centers offers cost-efficiency,
it requires robust isolation techniques to prevent unauthorized
access to tenant's sensitive data. 
The lack of fine-grained and dynamic access control
(such as Discretionary Access Control (DAC),
Mandatory Access Control (MAC), or combinations thereof), adequate
resource isolation, and information flow-control measures  
may allow attackers ($\mathcal{A}_3$ and $\mathcal{A}_4$) to obtain unauthorized
access to sensitive sustainability data, potentially leading to data breaches, privacy
violations, and other security issues.
Furthermore, sustainability data can also be illegitimately
tampered with by malicious users processes ($\mathcal{A}_6$) or
compromised system processes ($\mathcal{A}_5$).
Malicious processes may
obtain unauthorized (read/write) access to sensitive resources (\eg
databases or protected memory regions storing sustainability data and
states) by exploiting vulnerabilities in the access control
policies\footnote{\href{https://threatpost.com/bug-linux-kernel-privilege-escalation-container-escape/178808/}{https://threatpost.com/bug-linux-kernel-privilege-escalation-container-escape/178808/}}. 
%~\cite{privilege-escalation}.
As a result, regular sustainability
operations are likely to be disrupted, which may cause
the system to produce unwarranted carbon footprints, including a neutral footprint.
Tampering with sustainability data by attackers (\eg malicious
service providers or malicious users) may result in overcharging
legitimate users of the system (such as a data center), undercharging
malicious users attempting to evade sustainability costs, or damaging
the reputation of competing service providers.
Attackers may also induce
carbon-exhaustion attacks on other users by misreporting of carbon consumption
or evade compliance checking of regulatory agencies by misreporting
low carbon emissions when operating in test mode (similar to Volkswagen's
scandal\footnote{\href{https://www.cpajournal.com/2019/07/22/9187/}{https://www.cpajournal.com/2019/07/22/9187/}}).
%~\cite{vwscandal2015}).

\noindent \ding{113} \textbf{Sensitive information disclosure (C4).}
Collecting fine-grained sustainability data from disparate carbon sources
(\eg sensors and PDUs) to monitor and diagnose sustainability activities may also disclose
the sustainability metrics to service providers ($\mathcal{A}_2$) and other users.
Such unauthorized exposure of footprint will violate the privacy of users'
data, location, behavior, and intellectual properties such as
proprietary scheduling techniques, factors used for competitive
pricing for different service categories~\cite{hlavacs2011energy, mckenna2012smart}.
%\ezk{missing ``CITE''?}
Unauthorized access to footprint data
can enable an adversary to initiate DoS attacks ($\mathcal{A}_4$ and $\mathcal{A}_6$)
on the co-tenant and thus prevent co-tenants from realizing an
desired sustainability target.

\noindent \ding{113} \textbf{Cryptographic flaws and software bugs (C5).}
The ability of a sustainable system to provide proof of
carbon footprint to users and regulators is essential for ensuring the
trustworthiness of the system.  Such proof of footprint should be built
with cryptographic constructs.  However, flaws in the integration of
cryptographic constructs with complex data center systems (\eg using
weak cipher suites\footnote{\href{https://threatpost.com/bug-linux-kernel-privilege-escalation-container-escape/178808/}{https://threatpost.com/bug-linux-kernel-privilege-escalation-container-escape/178808/}}\footnote{\href{https://nakedsecurity.sophos.com/2023/01/30/serious-security-the-samba-logon-bug-caused-by-outdated-crypto/}{https://nakedsecurity.sophos.com/2023/01/30/serious-security-the-samba-logon-bug-caused-by-outdated-crypto/}})
%~\cite{ms365-insecure-block-cipher})
%~\cite{ms365-insecure-block-cipher,samba-outdated-crypto})
or flaws in the
software\footnote{\href{https://heartbleed.com/}{https://heartbleed.com/}}
%~\cite{heartbleed}
($\mathcal{A}_5$)
may fail to generate unforgeable and
accurate proof of consumption, enabling an attacker to drop, modify, replay,
and inject fake footprints of carbon.  This can disrupt the
operations of sustainable systems.

\noindent \ding{113} \textbf{Side-channels in sustainability (C6).}
Due to shared hardware resources, co-located tenants' servers, and poor
isolation between different processes running on the same hardware in
data centers, side-channel vulnerabilities~\cite{cloud-side-channel-ristenpart} %~\cite{harnik2010side}
(\eg page faults, %~\cite{xu2015controlled},
cache misses, %~\cite{wang2007new},
power, %~\cite{randolph2020power}
and timing %~\cite{hund2013practical}
channels) may allow a malicious
process ($\mathcal{A}_4$ and $\mathcal{A}_6$) to observe or tamper with carbon footprint
patterns of other users' jobs/applications running on the same
hardware.  Such side channels allow an attacker to not only
fingerprint the data traffic of other users but also extract the
cryptographic keys or other confidential information of a user
application by looking at the use of sustainability
metrics~\cite{cloud-side-channel-ristenpart}.
Attackers ($\mathcal{A}_3$ and $\mathcal{A}_5$) can exploit
such sensitive information to blackmail or embarrass other
users/competitors (\eg to force a competitor's stock to drop, or
short-sell such stock).

\noindent \ding{113} \textbf{Collusion for evasion (C7).}  Infrastructure
providers ($\mathcal{A}_7$) and Power Distribution Unit (PDU)
providers ($\mathcal{A}_3$) may collude to
misreport carbon footprints to regulators and users and thus may evade
regulatory agencies.
Such collusion attacks can be of different
combinations as infrastructure providers depend on third-party
software and hardware vendors which may also collude with each other
for malicious purposes.

\mycolor{black}

%%%%%%%%%%%%%%%%%%%%%%%%%%%%%%%%%%%%%%%%%%%%%%%%%%%%%%%%%%%%%%%%%%%%%%%%%%%%%%
%% For Emacs:
% Local variables:
% fill-column: 70
% End:
%%%%%%%%%%%%%%%%%%%%%%%%%%%%%%%%%%%%%%%%%%%%%%%%%%%%%%%%%%%%%%%%%%%%%%%%%%%%%%
%% For vim:
% vim:textwidth=70
%%%%%%%%%%%%%%%%%%%%%%%%%%%%%%%%%%%%%%%%%%%%%%%%%%%%%%%%%%%%%%%%%%%%%%%%%%%%%%
% LocalWords:  HotCarbon externalities Pigovian Ent Jevons Pigou TODO
% LocalWords:  unforgeable tesla youtube de facto SaaS PaaS IaaS RECs
% LocalWords:  PPAs PPA PDUs PDU

%% file: table-challenges.tex
\begin{table*}
\footnotesize
\centering
{\mycolor{blue}\begin{tabular}{|m{0.3cm}|m{2.7cm} | m{5.8cm} | m{3cm} | m{2.1cm}|}
\hline
\centering\textbf{ID} &
\centering\textbf{Threat Model} &
\centering\textbf{Vulnerabilities and Security Challenges} &
\centering\textbf{Impacts} &
\centering\textbf{Possible Ideas to Solutions}
\tabularnewline \hline

%\ding{187} &
\textbf{C1} &
\emph{Untrusted:} Data center operators ($\mathcal{A}_1$),
\emph{Trusted:} Other stakeholders
&
Evasive carbon offset techniques and lack of authenticity, accountability and transparency allow data center providers ($\mathcal{A}_1$)
to trade a known amount of carbon emissions with an uncertain
amount of carbon reductions
&
Tax evasion, financial loss, and environmental
hazards
&
Verifiable footprint collection (\S\ref{subsec:verifiable_collection})
\tabularnewline \hline

%\ding{182} &
\textbf{C2} &
\emph{Untrusted:} Data center operators ($\mathcal{A}_3$), tenants ($\mathcal{A}_4$), external attackers ($\mathcal{A}_5$),
\emph{Trusted:} Other stakeholders
&
The lack of integrity (tamper-proof guarantee) of carbon emission sources
allows malicious providers ($\mathcal{A}_3$), physically co-located tenants ($\mathcal{A}_4$) or
external attackers ($\mathcal{A}_5$) to forge,
tamper, or misreport carbon usage~\cite{shumailov2021sponge}
&
Cause over/under-billing to customers by tampering with carbon usage,
evade regulatory agencies by misreporting low carbon emissions
&
Verifiable footprint collection (\S\ref{subsec:verifiable_collection})
\tabularnewline \hline

%\ding{183} &
% \textbf{C3} &
% Untrustworthy physical environment may allow attackers to manipulate
% sensors and apparatuses within a data center directly or indirectly
% &
% Induce higher operational costs, cause
% over-/under-billing to customers, and denial-of-service attacks
% &
% Verifiable footprint collection (\S\ref{subsec:verifiable_collection})
% \tabularnewline \hline
%\ding{185} &
\textbf{C3} &
\emph{Untrusted:} Data center operators ($\mathcal{A}_3$), tenants ($\mathcal{A}_4$), external attackers ($\mathcal{A}_5$), users ($\mathcal{A}_6$),
\emph{Trusted:} Other stakeholders
&
Inadequate access control or information flow control
mechanisms may allow
attackers ($\mathcal{A}_3$ - $\mathcal{A}_6$) to
access and tamper with the databases storing carbon footprint trails
&
Exposure of users' private data such as location, behavior, and intellectual
properties
&
Verifiable footprint collection (\S\ref{subsec:verifiable_collection})
\tabularnewline \hline

%\ding{184} &
\textbf{C4 \& C6} &
\emph{Untrusted:} Data center operators ($\mathcal{A}_2$), tenants ($\mathcal{A}_4$), users ($\mathcal{A}_6$),
\emph{Trusted:} Other stakeholders
&
Disclosure of sustainability metrics to honest but curious $\mathcal{A}_2$ or malicious service providers ($\mathcal{A}_4$) and users ($\mathcal{A}_6$) due to inadequate access control,
cryptographic protections, or side-channel vulnerabilities
&
Exposure of users' private data such as location, behavior, and
intellectual properties
&
Privacy-preserving footprint collection and aggregation
(\S\ref{subsec:privacy-preserving-collection},
\S\ref{subsec:privacy-preserving-aggregation}, \&
\S\ref{subsec:public-sustainability-ledger})
\tabularnewline \hline

%\ding{186} &

\textbf{C5} &
\emph{Untrusted:} Data center operators ($\mathcal{A}_3$), tenants ($\mathcal{A}_4$), external attackers ($\mathcal{A}_5$), users ($\mathcal{A}_6$),
\emph{Trusted:} Other stakeholders
&
Cryptographic flaws and software vulnerabilities may
allow attackers ($\mathcal{A}_5$) to forge the proof of carbon usage
&
Financial loss and disruption the data center operations
&
Verifiable footprint collection (\S\ref{subsec:verifiable_collection})
\tabularnewline \hline

%\ding{188} &
\textbf{C7} &
\emph{Untrusted:} Data center operators ($\mathcal{A}_1$), tenants ($\mathcal{A}_4$), regulators ($\mathcal{A}_7$), users ($\mathcal{A}_6$),
\emph{Trusted:} Government
&
Multiple parties such as providers ($\mathcal{A}_3$) and users ($\mathcal{A}_6$) or providers ($\mathcal{A}_3$) and regulators ($\mathcal{A}_7$) may collude to misreport carbon usage
&
Tax evasion, financial loss, and environmental
hazards
&
Verifiable footprint collection (\S\ref{subsec:verifiable_collection})
\tabularnewline \hline

\end{tabular}
}
\mycolor{black}
\caption{Threats and security challenges for the sustainability of
  data centers and potential research directions.
  %\ag{subsection numbers not showing up in table?} %\David{move C3 two rows up?}
  }

\label{tab:sustainability_threats}
\end{table*}

%%%%%%%%%%%%%%%%%%%%%%%%%%%%%%%%%%%%%%%%%%%%%%%%%%%%%%%%%%%%%%%%%%%%%%%%%%%%%%
%% For Emacs:
% Local variables:
% fill-column: 70
% End:
%%%%%%%%%%%%%%%%%%%%%%%%%%%%%%%%%%%%%%%%%%%%%%%%%%%%%%%%%%%%%%%%%%%%%%%%%%%%%%
%% For vim:
% vim:textwidth=70
%%%%%%%%%%%%%%%%%%%%%%%%%%%%%%%%%%%%%%%%%%%%%%%%%%%%%%%%%%%%%%%%%%%%%%%%%%%%%%
% LocalWords:  HotCarbon externalities Pigovian Ent Jevons Pigou TODO
% LocalWords:  unforgeable tesla youtube de facto SaaS PaaS IaaS

%% file: verifying.tex
\section{Research Directions for Securing Sustainable Data Centers}
\label{s:verify}

\begin{figure*}[t]
\centering
\includegraphics[width=\textwidth]{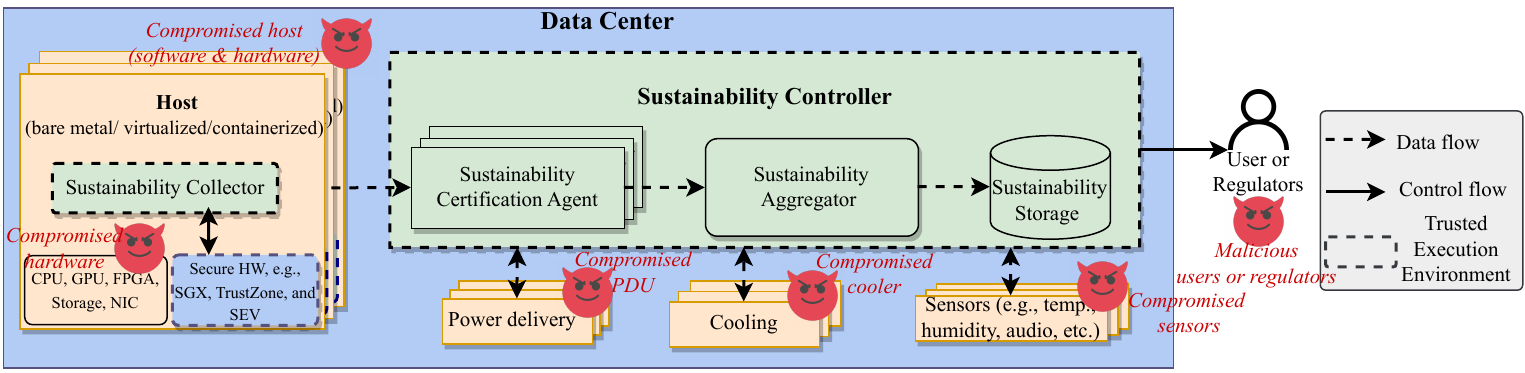}
\caption{To enable the verifiability of sustainability metrics, we
  propose that sustainability-aware data centers be equipped with a
  \emph{sustainability collector}, \emph{certification agent},
  \emph{sustainability aggregator}, and \emph{sustainability storage}.
  We mark components in the data center with an adversary symbol to
  denote potential compromised components. Unchanged items in data centers
  are shaded blue, modified items are shaded orange, and new items added for
  sustainability are in green.
  }
\label{fig:arch}
\end{figure*}

Although many solutions~\cite{confidential-computing}
%~\cite{berger2008tvdc, confidential-computing}
have been designed for data-center security, most of them are not directly applicable
to counter the security and privacy challenges towards sustainability as
discussed in Section~\ref{sec:security-challenges}.
Therefore, we must develop technologies that will help build
secure and trustworthy sustainable systems.
Particularly, we must develop primitives that allow domain experts to
construct and operate sustainable systems and verify the results.
Next, we lay out several potential research directions for improving
sustainability in data centers through security.

%%%%%%%%%%%%%%%%%%%%%%%%%%%%%%%%%%%%%%%%%%%%%%%%%%%%%%%%%%%%%%%%%%%%%%
\subsection{Verifiable Footprint Collection Architecture}
\label{subsec:verifiable_collection}

One of the most important elements of a sustainable system is its
ability to promote the responsible use of system resources, such as
complying with carbon emission restrictions/taxes.  However, claims of
carbon usage must be accompanied by infrastructure that demonstrates
\emph{verifiable footprint} to the public and regulatory organizations.
This calls for architectures and systems that can collect publicly
readable and verifiable sensor readings in adversarial settings.  It
is essential that these systems have the ability to scale seamlessly
from small, low-energy devices to larger, enterprise-level data
centers.  The system architecture should have the ability to generate
tamper-resistant proofs of carbon consumption that are unforgeable,
accurate, and securely retrievable by authorized parties (which might
include the public) in adversarial deployments.  Furthermore, to provide higher
security assurance, the design and implementation of these systems
must be formally verified.

\noindent \textbf{Potential Solutions}:
\mycolor{blue}
Developing such a framework poses key challenges, including the need
to establish and preserve a root of trust using trusted hardware, such
as Trusted Platform Module (TPM)
to secure the data center's carbon footprint measurement components.
A trusted path should be established from the secure hardware up to the
module that collects all the relevant metrics of a job, and further up
to the component that verifies the accuracy of the reported
metrics.
\mycolor{black}
This trusted path will be capable of producing tamper-proof
evidence of sustainability cost metrics using cryptographic proof
systems.

One potential solution to ensure the security of
sustainability-related components is to use a hardware-based Trusted
Execution Environment (TEE) such as ARM TrustZone, 
%\footnote{\href{https://developer.arm.com/documentation/100690/0201}{https://developer.arm.com/documentation/100690/0201}},
%~\cite{armtz2016},
Intel SGX, 
%\footnote{\href{https://www.intel.com/content/www/us/en/architecture-and-technology/software-guard-extensions.html}{https://www.intel.com/content/www/us/en/architecture-and-technology/software-guard-extensions.html}},
%~\cite{mckeen2013innovative},
AMD SEV, 
%\footnote{\href{https://www.amd.com/en/developer/sev.html}{https://www.amd.com/en/developer/sev.html}}
%~\cite{kaplan2016sev},
and Keystone. 
%\footnote{\href{https://github.com/keystone-enclave/keystone}{https://github.com/keystone-enclave/keystone}}.
%~\cite{lee2020keystone}.
%
%\David{I doubt we need these footnotes (maybe Keystone)}
TEEs are deployed in nearly every commercial processor sold today and
are the de-facto standard to provide a tamper-proof execution
environment that preserves the integrity and confidentiality of data
and execution~\cite{graphene_sgx_atc17}.
% , scone_osdi16,
%   sgx_fv_usenix21}.
%
These environments provide isolation guarantees needed to certify that
metric data is collected and reported accurately, even in the presence
of malicious applications, OS, or hypervisor.
A \emph{sustainability collector} (see Figure~\ref{fig:arch}) running
in a TEE will securely collect the utilization details of a
bare-metal, virtualized, or containerized job.
The gathered metrics
will create a comprehensive timeline of user, system, and
process-oriented carbon footprints, culminating in a
\emph{sustainability provenance record} for the cloud.
The sustainability collector will securely report the metrics to a
\emph{sustainability certification agent}, which will produce
lightweight cryptographic proofs that empower third-party regulators
and users to independently verify the claimed consumption.

\mycolor{blue}
Note that any flaws in the design or implementation of
sustainability-related components, \eg measurement or collection code
running within TEEs and owned by respective TEE hosting entities
(\ie data center operators or service providers)
may introduce new security challenges.  For instance,
attackers may exploit such flaws and bypass the tamper-proof guarantees
of the code.
Therefore, it is crucial to ensure high-security assurance of these components through formal
analysis before they are deployed.
Also, the physical or virtual machines hosting the measurement code within TEEs
and the regulatory agencies need to verify during runtime the integrity of the trusted
path from the secure hardware to corresponding TEEs periodically
or when there are major changes (\eg write operations) in the system or
their combinations thereof.
\mycolor{black}

Another potential concern is that current TEE platforms
might lack adequate privileges to monitor the carbon or resource consumption
of workloads that execute outside of the TEE.  This might necessitate
new hardware support for TEEs to allow secure monitoring of external
workloads, including the host OS or hypervisor.

One possible alternative to TEEs is to explore the use of add-on
monitoring hardware, akin to SmartNICs, that can collect
sustainability metrics from outside the host.
For example, AWS Nitro\footnote{\href{https://aws.amazon.com/ec2/nitro/}{https://aws.amazon.com/ec2/nitro/}}
%~\cite{nitro}
enables SmartNICs to monitor and
manage VM allocation and scheduling, while being technically
``outside'' the host OS.
Similarly, sustainability-related components could potentially run on
such add-on custom hardware with the necessary privileges to gather
data from the host without being vulnerable to compromise by the host.
Finally, sustainability data must be isolated from other workloads running on
the same machine, providing protection against unauthorized access and
tampering.

%%%%%%%%%%%%%%%%%%%%%%%%%%%%%%%%%%%%%%%%%%%%%%%%%%%%%%%%%%%%%%%%%%%%%%
\subsection{Privacy-Preserving Footprint Collection}
\label{subsec:privacy-preserving-collection}
Fine-grained sustainability data collected through disparate carbon sources, such
as sensors and PDUs in an unregulated manner, may induce unintended
disclosure of sensitive data.
The exposure of sustainability records would otherwise break the
users' privacy, data, location, behavior, and intellectual properties
such as proprietary scheduling techniques, trained machine learning
models, and factors used for competitive pricing for service classes~\cite{hlavacs2011energy, mckenna2012smart}.
Also, attackers may attempt to tamper with sensor data before it
is aggregated, which can lead to incorrect or misleading results. This
can be especially problematic in safety-critical applications, such as
autonomous vehicles or medical devices.

\noindent \textbf{Potential Solutions}: In concert with the verifiable
sustainability data collection architecture, differential privacy (DP) or
local differential privacy (LDP)
can be used as a probabilistic solution for privacy-preserving sustainability
footprint collection.
A certain degree of noise can be added to
the collected data to obscure individual data points but still allow
for useful aggregate analysis~\cite{dwork2006differential}.
%, dwork2008differential}.
%
\mycolor{blue}
A classical challenge of such differential privacy-based solutions would
be to keep the utility (\eg the statistical properties) of the sustainability data
high to the system while still protecting the privacy of users and systems.
In other words, the privacy budget---the amount of noise that can be
added to the sustainability data without compromising privacy---needs
to be determined by the sensitivity of the sustainability data being
collected and the desired level of privacy protection.
Another challenge for DP-based solutions is to keep the total noise
added by all parties within an acceptable range and failure to do so
requires a trusted aggregator to correct the noise.
Since DP-based solutions protect the data owner by providing
indistinguishability of the dataset, they can be used as a
privacy-preserving way of releasing data.  However, one has to ensure
correct-by-constructions~\cite{morinov2012ccs, jin2022we} of such while adopting
them.
\mycolor{black}

\mycolor{blue}
To provide cryptographic guarantees and to preserve the utility of sustainability data
utility to a higher extent compared to differential privacy,
an alternate solution is to use homomorphic encryption\footnote{Craig Gentry. A fully homomorphic encryption
scheme. Ph.D. Dissertation, Stanford university, 2009.}. %~\cite{gentry2009fully}.
With this solution, the carbon sources can encrypt the sustainability data as well as enable
the decision-making agent to measure/compute any statistical
information on those encrypted data.
\mycolor{black}
There are, however, several challenges associated with this
solution. 
%~\cite{gentry2009fully}.
%~\cite{naehrig2011can}.
%
%\David{I doubt we need this footnote}
Homomorphic encryption (HE) requires significant computational resources
and can increase the size of the actual data (because of encryption) being
transmitted,
%~\cite{naehrig2011can},
making it more difficult to store and transmit efficiently.
Furthermore, there are currently limitations
%~\cite{naehrig2011can}
on the types of computations that can be performed on homomorphically
encrypted data.  For example, homomorphic encryption schemes support
only addition and multiplication.  Complex operations, such as
division or trigonometric functions, may not be efficiently supported. 
\mycolor{blue}
While the direct use of homomorphic encryption may not be appropriate
for resource-constrained carbon emission sources,
further research is warranted to check if optimized versions of HE such as
partial HE, leveled HE, and threshold HE can be utilized or a
new, lightweight, secure, and bespoke HE (\eg selective HE)
needs to be designed for sustainability in data centers.
Nevertheless, many major chip/system vendors such as
Intel, AMD and ARM are actively exploring hardware support for
HE and when these are available, they can provide a trusted basis for
implementing challenges to many of the security solutions identified in this article.
\mycolor{black}

Another alternative approach
involving less computational overhead than homomorphic encryption is
zero-knowledge proofs~\cite{zero-knowledge-proof},
%~\cite{fiege1987zero},
in which the carbon sources
can demonstrate to the sustainability certification agent,
that sustainability footprints are valid, without disclosing
the actual values that would otherwise compromise privacy.
\mycolor{blue}
However, zero-knowledge proofs can only be used to prove the authenticity of sustainability data
and is not intended for analyzing and making any decisions.
To address the challenges of each solution, further investigation is needed
to determine if homomorphic encryption or differential privacy can be
combined with zero-knowledge proofs.

\mycolor{black}

%%%%%%%%%%%%%%%%%%%%%%%%%%%%%%%%%%%%%%%%%%%%%%%%%%%%%%%%%%%%%%%%%%%%%%
\subsection{Privacy-Preserving Footprint Aggregation}
\label{subsec:privacy-preserving-aggregation}
Collecting and processing sustainability data from multiple sites in
data centers require secure collaboration between multiple untrusted
parties, including cloud operators, regulators, and users, each with
their own confidentiality, privacy, security, and trust requirements.
While being aggregated either in centralized or distributed data
centers, sustainability data can still reveal sensitive information
about users and systems as discussed in Section~\ref{subsec:new-security-challenges}.
Therefore, the high-level goals are to (1) perform aggregation,
summary, or other functions on the sustainability data whose results
do not disclose information about the underlying data; and (2) ensure
that aggregations provide (provably) accurate higher-level data
without exposing underlying sensitive information, \eg proof of
sustainability compliance of the manufacturing process without
exposing unit-wise behaviors or specific metrics.

\noindent \textbf{Potential Solutions}: A plausible approach to
privacy-preserving aggregation for sustainability data is
secure multi-party communication (MPC) in which multiple
carbon footprint aggregators located at different locations
collaborate to perform computations on their combined data without
revealing any individual data points~\cite{goldreich1998secure}.
MPC requires minimal trust and aims to ensure each party's
input is kept private while allowing them to compute the
desired aggregation, summary, or other
functions on their combined data whose results do not disclose
information about the underlying data.
One such MPC platform is Confidential Space by Google\footnote{\href{https://cloud.google.com/blog/products/identity-security/announcing-confidential-space}{https://cloud.google.com/blog/products/identity-security/announcing-confidential-space}}, 
%~\cite{confidential-space},
which would allow sustainability data to be encrypted and stored in a
TEE that only authorized workloads are allowed to access.
Additionally, such data is isolated from other workloads running on
the same machine, protecting unauthorized access and
tampering.
\mycolor{blue}
MPC-based solutions, however, incur higher computational and communication
overheads due to secure computations and sharing of encrypted results.
%To prevent an adversary from linking a particular record
%to a specific party.An alternate solution is to use K-anonymity

%

%\David{The footnote for FL is not needed}

To minimize sustainability data movement, federated learning 
%\footnote{\href{https://blog.research.google/2017/04/federated-learning-collaborative.html}{https://blog.research.google/2017/04/federated-learning-collaborative.html}}
%\cite{li2020federated}
can be used in which training a machine learning model (\eg carbon footprint
optimization) on decentralized sustainability data/metrics can be
performed without having to transfer the data to a centralized
location.
\mycolor{black}
Each site of the distributed data center will train a local model on
its sustainability data and send the updated model weights to a
central server, which aggregates them to create a global model.
This approach allows data to remain local and private while still
benefiting from a centralized learning process.
Note that existing federated learning techniques are susceptible to
model-poisoning and model-stealing attacks; this further imposes
challenges to adopt federated learning-based solutions for aggregating
sustainability data. %~\cite{li2020federated}.

%%%%%%%%%%%%%%%%%%%%%%%%%%%%%%%%%%%%%%%%%%%%%%%%%%%%%%%%%%%%%%%%%%%%%%
\subsection{Public Sustainability Ledgers}
\label{subsec:public-sustainability-ledger}

Public sustainability ledgers can be used for tracking carbon
emissions or energy consumption and thus can provide transparency and
accountability in the management of resources.
However, there are also security and privacy issues that need to be
considered when using these public ledgers.
For example, if public ledgers contain sensitive data (\eg carbon
credit allocations, sales, and expenditures) about the sustainability
practices of individuals and organizations, attackers may track the
individuals/organizations or infer proprietary algorithms.
Also, sustainability data may be stored on multiple public ledgers or
private databases, which may not be interoperable.
This can create challenges in ensuring data consistency and accuracy,
and may also lead to data breaches if not properly secured.

\noindent \textbf{Potential Solutions}: In combination with
privacy-preserving measures, such as homomorphic encryption,
zero-knowledge proofs, multi-party computations, and differential
privacy, public ledgers for sustainability reporting can be provided
through smart contracts~\cite{aloqaily2020energy} deployed on the
public blockchain.
The smart contract records the sustainability footprints from different sources
and stores the encrypted records in blocks on the blockchain.
The sustainability footprints submitted to the blockchain undergo
verification by the participating entities through a consensus mechanism,
such as Proof-of-Work (PoW) or Proof-of-Stake (PoS).
This ensures the accuracy
and integrity of the recorded footprints.
Consumers, stakeholders, and regulators can access the public blockchain
to track and verify the provenance of sustainability footprints.
Although smart contracts---in concert with a verifiable
sustainability footprint collection architecture (Figure~\ref{fig:arch}) and
privacy-preserving measures---can offer secure and public sustainability ledgers,
smart contracts can also be subject to vulnerabilities that can be exploited by
attackers. %~\cite{perez2021smart}.
As such, it is important to thoroughly test and audit smart contracts
to ensure their security and reliability. %~\cite{tsankov2018securify}.
Furthermore, blockchain technology 
%\footnote{\href{https://www.ibm.com/topics/blockchain}{https://www.ibm.com/topics/blockchain}}
%~\cite{pilkington2016blockchain}
can be used to address the inconsistency and data-breach issues of
distributed public ledgers.
However, current blockchain technologies are susceptible to various
types of attacks including 51\% (majority) attacks and
denial-of-service attacks. %~\cite{zhang2019security}.
As such, it is important to ensure that the blockchain network is
properly secured and appropriate security measures are in place
to prevent such attacks.

\mycolor{blue}
While the potential security solutions outlined in this paper may contribute to carbon footprints, future research is necessary to rigorously evaluate the performance
and security guarantees of the existing and newly designed solutions.
As discussed in Section~\ref{sec:security-challenges}, the importance
of such security solutions in ensuring the trustworthiness of sustainability data and incentivizing the users toward sustainability practices is crucial for addressing global climate change and is believed to outweigh the impact of systems lacking such guarantees.

\mycolor{black}

\mycolor{blue}
%%%%%%%%%%%%%%%%%%%%%%%%%%%%%%%%%%%%%%%%%%%%%%%%%%%%%%%%%%%%%%%%%%%%%%
\section{Enhancing Standardization of Security
Mechanisms}

Security mechanisms are essential to ensure
compliance with regulations and standards,
preventing unauthorized access, and exposure,
tampering, or misuse of sustainability data.
Irrespective of the specific solution used to ensure security
of sustainability, a common need is to ease the adoption of those mechanisms
and reduce their footprint, both in terms of performance and
sustainability.
For instance, a TEE-based solution for verifiable data collection or
a homomorphic encryption-based approach for privacy-preserving
footprint collection should be lightweight and have small footprints
so as to minimize overall carbon consumption.
As trustworthiness is foundational in sustainability initiatives,
stakeholders, including governments, businesses, and users,
need high security and privacy assurance of sustainability data, which
is crucial for the success and adoption of sustainability practices.
Since sustainability data is critical for understanding trends
and for long-term planning and monitoring to counter global
issues such as climate change,
it is necessary to rigorously evaluate the effectiveness of security measures toward sustainability to
make informed decisions for a sustainable future.
As sustainability is a global concern that requires collaboration across borders,
standardizing security mechanisms for sustainability data
will accelerate their adoption in other sectors, facilitate international
cooperation, and ensure consistent protection standards and interoperability. 
Incentives and regulations need to be introduced to motivate
organizations to adopt and implement standardized security
mechanisms.
These could include tax incentives, certification programs, or
regulatory requirements that prioritize sustainability and security.
Note that all challenges toward sustainability
cannot be solved with technical solutions alone.
Hence, offering both fundamental principles and secure guarantees is more likely to assist
in the development of policies.  This, in turn, can contribute to and accelerate the global effort to combat climate change.
Without robust policies, all optimizations are susceptible to the
Jevons paradox (i.e., increasing efficiency can lead to 
increased consumption), which signifies that both regulation and security
are crucial components. Hence, collaboration and cooperation among industry players, researchers, and policymakers are necessary to establish these common goals and objectives.
\mycolor{black}
%\zhenhua{Jevons paradox perhaps needs some explanation.}

%%%%%%%%%%%%%%%%%%%%%%%%%%%%%%%%%%%%%%%%%%%%%%%%%%%%%%%%%%%%%%%%%%%%%%%%%%%%%%
%% For Emacs:
% Local variables:
% fill-column: 70
% End:
%%%%%%%%%%%%%%%%%%%%%%%%%%%%%%%%%%%%%%%%%%%%%%%%%%%%%%%%%%%%%%%%%%%%%%%%%%%%%%
%% For vim:
% vim:textwidth=70
%%%%%%%%%%%%%%%%%%%%%%%%%%%%%%%%%%%%%%%%%%%%%%%%%%%%%%%%%%%%%%%%%%%%%%%%%%%%%%
% LocalWords:  PSU unforgeable incentivization TLS allocator SGX SEV
% LocalWords:  TrustZone de facto TEEs SASSY's compositional crypto
% LocalWords:  TEE's HSC sss TODO PDUs IPsec RAPL XMLERS Merkle LDP
% LocalWords:  cryptographically adaptively incentivized SmartNICs
% LocalWords:  PoW PoS Jevons

%% file: conc.tex
\section{Conclusion}
\label{s:conc}
Security infrastructure for a sustainable system is indispensable
for protecting the environment and our planet.
The central goal of this security infrastructure
is to enable service providers to produce unforgeable proofs of
sustainability footprints for users or
regulators while preventing potential security and privacy
threats by malicious users or compromised systems.
Towards this goal, this paper discusses the threat landscapes and
new security challenges to achieve sustainability of data centers and
presents potential
 research directions to develop primitives that allow domain experts to
construct and operate sustainable data centers.
The proposed challenges and potential solutions also lay the foundations
for other sustainable systems, such as manufacturing, telecommunication systems,
and automated transportation systems.

%%%%%%%%%%%%%%%%%%%%%%%%%%%%%%%%%%%%%%%%%%%%%%%%%%%%%%%%%%%%%%%%%%%%%%%%%%%%%%
%% For Emacs:
% Local variables:
% fill-column: 70
% End:
%%%%%%%%%%%%%%%%%%%%%%%%%%%%%%%%%%%%%%%%%%%%%%%%%%%%%%%%%%%%%%%%%%%%%%%%%%%%%%
%% For vim:
% vim:textwidth=70
%%%%%%%%%%%%%%%%%%%%%%%%%%%%%%%%%%%%%%%%%%%%%%%%%%%%%%%%%%%%%%%%%%%%%%%%%%%%%%
% LocalWords:  PSU unforgeable

%% file: main.bbl
\begin{thebibliography}{10}

\bibitem{aloqaily2020energy}
Moayad Aloqaily, Azzedine Boukerche, Ouns Bouachir, Fariea Khalid, and Sobia
  Jangsher.
\newblock An energy trade framework using smart contracts: Overview and
  challenges.
\newblock {\em IEEE Network}, 34(4):119--125, 2020.

\bibitem{graphene_sgx_atc17}
Chia che Tsai, Donald~E. Porter, and Mona Vij.
\newblock {Graphene-SGX}: A practical library {OS} for unmodified applications
  on {SGX}.
\newblock In {\em 2017 USENIX Annual Technical Conference (USENIX ATC 17)},
  pages 645--658, Santa Clara, CA, July 2017. USENIX Association.

\bibitem{dwork2006differential}
Cynthia Dwork.
\newblock Differential privacy.
\newblock In {\em Automata, Languages and Programming: 33rd International
  Colloquium, ICALP 2006, Venice, Italy, July 10-14, 2006, Proceedings, Part II
  33}, pages 1--12. Springer, 2006.

\bibitem{gandhi2022metrics}
Anshul Gandhi, Kanad Ghose, Kartik Gopalan, Syed~Rafiul Hussain, Dongyoon Lee,
  David Liu, Zhenhua Liu, Patrick McDaniel, Shuai Mu, and Erez Zadok10.
\newblock Metrics for sustainability in data centers.
\newblock In {\em Proceedings of the 1st Workshop on Sustainable Computer
  Systems Design and Implementation (HotCarbon'22)}, 2022.

\bibitem{goldreich1998secure}
Oded Goldreich.
\newblock Secure multi-party computation.
\newblock {\em Manuscript. Preliminary version}, 78(110), 1998.

\bibitem{zero-knowledge-proof}
S~Goldwasser, S~Micali, and C~Rackoff.
\newblock The knowledge complexity of interactive proof-systems.
\newblock In {\em Proceedings of the Seventeenth Annual ACM Symposium on Theory
  of Computing}, STOC '85, New York, NY, USA, 1985. Association for Computing
  Machinery.

\bibitem{hlavacs2011energy}
Helmut Hlavacs, Thomas Treutner, Jean-Patrick Gelas, Laurent Lefevre, and
  Anne-Cecile Orgerie.
\newblock Energy consumption side-channel attack at virtual machines in a
  cloud.
\newblock In {\em 2011 IEEE Ninth International Conference on Dependable,
  Autonomic and Secure Computing}, pages 605--612. IEEE, 2011.

\bibitem{jin2022we}
Jiankai Jin, Eleanor McMurtry, Benjamin~IP Rubinstein, and Olga Ohrimenko.
\newblock Are we there yet? {Timing} and floating-point attacks on differential
  privacy systems.
\newblock In {\em 2022 IEEE Symposium on Security and Privacy (SP)}, pages
  473--488. IEEE, 2022.

\bibitem{masanet20}
Eric Masanet, Arman Shehabi, Nuoa Lei, Sarah Smith, and Jonathan Koomey.
\newblock Recalibrating global data center energy-use estimates.
\newblock {\em Science}, 367(6481):984--986, 2020.

\bibitem{mckenna2012smart}
Eoghan McKenna, Ian Richardson, and Murray Thomson.
\newblock Smart meter data: Balancing consumer privacy concerns with legitimate
  applications.
\newblock {\em Energy Policy}, 41:807--814, 2012.

\bibitem{morinov2012ccs}
Ilya Mironov.
\newblock On significance of the least significant bits for differential
  privacy.
\newblock In {\em Proceedings of the 2012 ACM Conference on Computer and
  Communications Security}. Association for Computing Machinery, 2012.

\bibitem{ren12}
Chuangang Ren, Di~Wang, Bhuvan Urgaonkar, and Anand Sivasubramaniam.
\newblock {Carbon-Aware Energy Capacity Planning for Datacenters}.
\newblock In {\em {Proceedings of the 20th IEEE International Symposium on
  Modeling, Analysis and Simulation of Computer and Telecommunication
  Systems}}, MASCOTS'12, pages 391--400, 2012.

\bibitem{cloud-side-channel-ristenpart}
Thomas Ristenpart, Eran Tromer, Hovav Shacham, and Stefan Savage.
\newblock Hey, you, get off of my cloud: Exploring information leakage in
  third-party compute clouds.
\newblock In {\em Proceedings of the 16th ACM Conference on Computer and
  Communications Security}, CCS '09, pages 199---212, New York, NY, USA, 2009.
  Association for Computing Machinery.

\bibitem{shumailov2021sponge}
Ilia Shumailov, Yiren Zhao, Daniel Bates, Nicolas Papernot, Robert Mullins, and
  Ross Anderson.
\newblock Sponge examples: Energy-latency attacks on neural networks.
\newblock In {\em 2021 IEEE European symposium on security and privacy
  (EuroS\&P)}, pages 212--231. IEEE, 2021.

\bibitem{confidential-computing}
Jianping Zhu, Rui Hou, XiaoFeng Wang, Wenhao Wang, Jiangfeng Cao, Boyan Zhao,
  Zhongpu Wang, Yuhui Zhang, Jiameng Ying, Lixin Zhang, and Dan Meng.
\newblock Enabling rack-scale confidential computing using heterogeneous
  trusted execution environment.
\newblock In {\em 2020 IEEE Symposium on Security and Privacy (SP)}, pages
  1450--1465, 2020.

\end{thebibliography}
